\pdfoutput=1
\documentclass[12pt]{iopart}
\usepackage[utf8]{inputenc}
\usepackage{iopams}
\usepackage{amsfonts}
\usepackage{graphicx}
\usepackage{upgreek}

\usepackage[table]{xcolor}

\newcommand*\rfrac[2]{{}^{#1}\!/_{#2}}
\newcommand*\cc{\cellcolor{gray!45}}   

\begin{document}
\maketitle
\title[Improving absolute gravity estimates by the $L_p$-norm approximation]{Improving absolute gravity estimates by the $L_p$-norm approximation of  the ballistic trajectory}
\author{V D Nagornyi}
\address{Metromatix, Inc., 111B Baden Pl, Staten Island, NY 10306, USA}
\ead{vn2@member.ams.org}

\author{S Svitlov}
\address{Institut f\"ur Erdmessung, Leibniz Universit\"at Hannover, Schneiderberg 50, D-30167 Hanover, Germany} 
\ead{svitlov@ife.uni-hannover.de}
\author{A Araya}
\address{Earthquake Research Institute, University of Tokyo, 1-1-1, Yayoi, Bunkyo-ku, Tokyo 113-0032, Japan} 
\ead{araya@eri.u-tokyo.ac.jp}
\begin{abstract}
Iteratively Re-weighted Least Squares (IRLS) were used to simulate the $L_p$-norm approximation of the ballistic trajectory in absolute gravimeters. Two iterations of the IRLS delivered sufficient accuracy of the approximation without a significant bias. The simulations were performed on different samplings and perturbations of the trajectory.
For the platykurtic distributions of the perturbations, the $L_p$-approximation with $3<p<4$ was found to yield several times more precise gravity estimates compared to the standard least-squares. The simulation results were confirmed by processing real gravity observations  performed at the excessive noise conditions. 
\end{abstract}
\section{Introduction}
Absolute ballistic gravimeters measure gravity acceleration by tracking the free motion of the test mass in the gravity field. Having the distances $\{S_1, ..., S_N\}$ covered by the test mass over the time intervals $\{T_1, ..., T_N\}$, the acceleration 
is found as parameter of some trajectory model fitted to the data pairs $(T_i, S_i)$ by methods of the regression analysis. One of the commonly used models describes unperturbed vertical motion of the test mass in the gravity field  
\begin{equation}
\label{eq_z_i}
z_i = z_0 + V_0\, T_i + g\, T_i^2/2.
\end{equation}
The expression (\ref{eq_z_i}) represents a linear regression model, because it's defined by a linear combination of the estimated parameters $z_0, V_0, g$. The estimates are usually found by minimising the sum of the squared discrepancies between the measured data and the model:
\begin{equation}
\label{eq_OLS}
(z_0, V_0, g): \;\; \sum  \epsilon_i^2 \rightarrow  \textrm{min}, \;\;\; \textrm{where} \;\;\; \epsilon_i = S_i - z_i.
\end{equation}
Minimised here is the square norm ($L_2$-norm) of the vector $\bepsilon=\{\epsilon_i\}$, which is a special case of the more general $L_p$-norm
\begin{equation}
\label{eq_p_norm}
\|\bepsilon\|_p = \left(\sum |\epsilon_i|^p\right)^{\rfrac{1}{p}}, \; p\ge 1.
\end{equation}
The tradition of minimising the sum of squares (i.e. finding the \emph{least squares}) has a long and reach history dating back to the beginning of the 19-th century. At that time the main reason for using the least squares was their computational simplicity, as for the $p$=2 the fitting of a linear model is reduced to finding a unique solution for a full-rank system of linear equations.  At any other $p$ the equations become non-linear in estimated parameters and require involved computations to find the solution.

Another reason for using the least squares relates to a different kind of linearity: linearity of the estimates with respect to the dependent variable ($S_i$ in our case), which implies no bias of the estimates. 
Moreover, according to the Gauss-Markov theorem, of all the linear/unbiased estimates, the least-squares deliver the most precise one, in case the errors in the observations are uncorrelated, have zero mean and constant variance. This is especially important in metrology, where the bias may lead to an incorrect estimates of the measurand and its uncertainty.

Nowadays, when advanced computing is no longer a problem, the $L_p$-approximation with $p \ne 2$ is widely used \cite{pennecchi2006}--\cite{ scales1988a}.
Many researches have noticed that the value of $p$ delivering the lowest variance to the $L_p$-estimates decreases as the kurtosis\footnote{The kurtosis $\upbeta_2$ characterises the spread of the distribution over the possible values of the random variable. It's defined as $\upmu_4/\upsigma^4$, where $\upmu_4$ is the fourth central moment, $\upsigma$ is the standard deviation.}  of the error distribution grows; there are formulae suggested for this dependence \cite{money1982, sposito1983}.
Most results have been obtained for the values $1 < p < 2$, that provide better estimates when the noises have heavily-tailed distributions.
In trajectory tracking, however, more prevalent are limited-band noises, for which the properties of the $L_p$-approximation are less known.

The purpose of this work is to investigate the feasibility and possible advantages of approximating the gravimeter's test mass trajectory in different $L_p$-norms. 

The paper has the following structure. The chapter \ref{sec_IRLS} discusses implementation of the Iteratively Re-weighted Least Squares (IRLS) for the $L_p$-norm approximation. The chapter \ref{sec_simulation} describes the Monte-Carlo simulation of gravity estimates in $L_p$-norm. A real-life example of improving the estimates by the $L_p$-norm approximation is discussed in the chapter \ref{sec_examples}. We summarise our study in the chapter \ref{sec_conclusions}.
\section{The $L_p$-norm approximation by the Iteratively Re-weighted least-squares}
\label{sec_IRLS}
We implement the $L_p$-norm approximation by performing the weighted least-squares (WLS) approximation, in which the weights are defined in a special way \cite{burrus2012}.
For the model (\ref{eq_z_i}), the WLS solution is determined by
\begin{equation}
\label{eq_WLS}
\bi{\bxi} = \bi{A}^+ \bi{S},
\end{equation}
where\\
$\bi{\bxi}=(z_0, V_0, g)^T$ is the vector of the unknown parameters,\\
$\bi{S}=(S_1, S_2, ..., S_N)^T$ is the vector of the observations,\\
$\bi{A}^+$ is the Moore-Penrose pseudo-inverse matrix calculated as
\begin{equation}
\label{eq_Aplus}
\bi{A}^+=(\bi{A}^T\bi{V}\bi{A})^{-1}\bi{A}^T\bi{V},
\end{equation}
where
\begin{equation}
\label{eq_A}
\bi{A} =
\left(
\begin{array}{ccc}
1       & T_1    & T_1^2/2 \\
1       & T_2    & T_2^2/2 \\
\vdots  & \vdots & \vdots  \\
1       & T_{N}  & T_{N}^2/2
\end{array}
\right)
\end{equation}
is the experiment design matrix,\\
\begin{equation}
\label{eq_V}
\bi{V}=\textrm{diag}(v_1, v_2, ..., v_N)
\end{equation}
is the diagonal matrix of weights.
The approximation in the $L_p$-norm is achieved when the weights $v_i$ are related to the discrepancies $\epsilon_i$ like \cite{burrus2012}
\begin{equation}
\label{eq_vi}
v_i = |\epsilon_i|^{p-2}.
\end{equation}
For $p$=2, the weights (\ref{eq_vi}) turn into units, yielding the standard least-squares solution.
The formula (\ref{eq_vi}) cannot be used directly to find the weights $v_i$, because the weights need to be known before the approximation, while the residuals $\epsilon_i$ are known only after the approximation is done. Because of that, the WLS is usually applied several times leading to the Iteratively Re-weighted Least Squares. The iterations start with the weights $v_i$=1, i.e. with the regular least-squares. The residuals $\epsilon_i$ are then used to build the new weights
\begin{equation}
\label{eq_V_by_e}
\bi{V}=\textrm{diag}(|\epsilon_1|^{p-2}, |\epsilon_2|^{p-2}, ..., |\epsilon_N|^{p-2}),
\end{equation}
which are then applied to approximate the original data. The iterations can be repeated, but the process is known to sometimes have a problematic convergence. Though different approaches exist to improve the convergence  \cite{burrus2012}, the simulations found that two iterations always produce a sufficiently accurate for our purposes approximation, with the bias not discernible in random noise.  
For better computational stability we normalised the weights (\ref{eq_vi}), so they would range from 0 to 1:
\begin{equation}
v_i = \left(\frac{|\epsilon_i|}{M} \right)^{p-2}, \;\;\; p \ge 2,
\end{equation}
where
\begin{equation}
M = \max|\epsilon_i|.
\end{equation}
If $p<2$, the powers in (\ref{eq_vi}) become negative, and to prevent zero division errors another modification of the weights was necessary:
\begin{equation}
\label{eq_threshold}
v_i = \left(\frac{\max (|\epsilon_i|, rM)}{rM}\right)^{p-2} ,\qquad p<2,
\end{equation}
where $r=0.001$ is a regularisation parameter establishing the threshold of smallness as fraction $r$ of the maximum residual $M$. According to (\ref{eq_threshold}), the residuals below the threshold get replaced with the threshold value $rM$. A similar way of regularisation is implemented in \cite{scales1988a}.
\section{Modelling the properties of the absolute gravity estimates in the $L_p$-norm}
\label{sec_simulation}
\subsection{Organisation of the simulation experiment}
\label{ssec_org_exp}
\begin{figure}[ht]
\centering
\includegraphics[height=100mm]{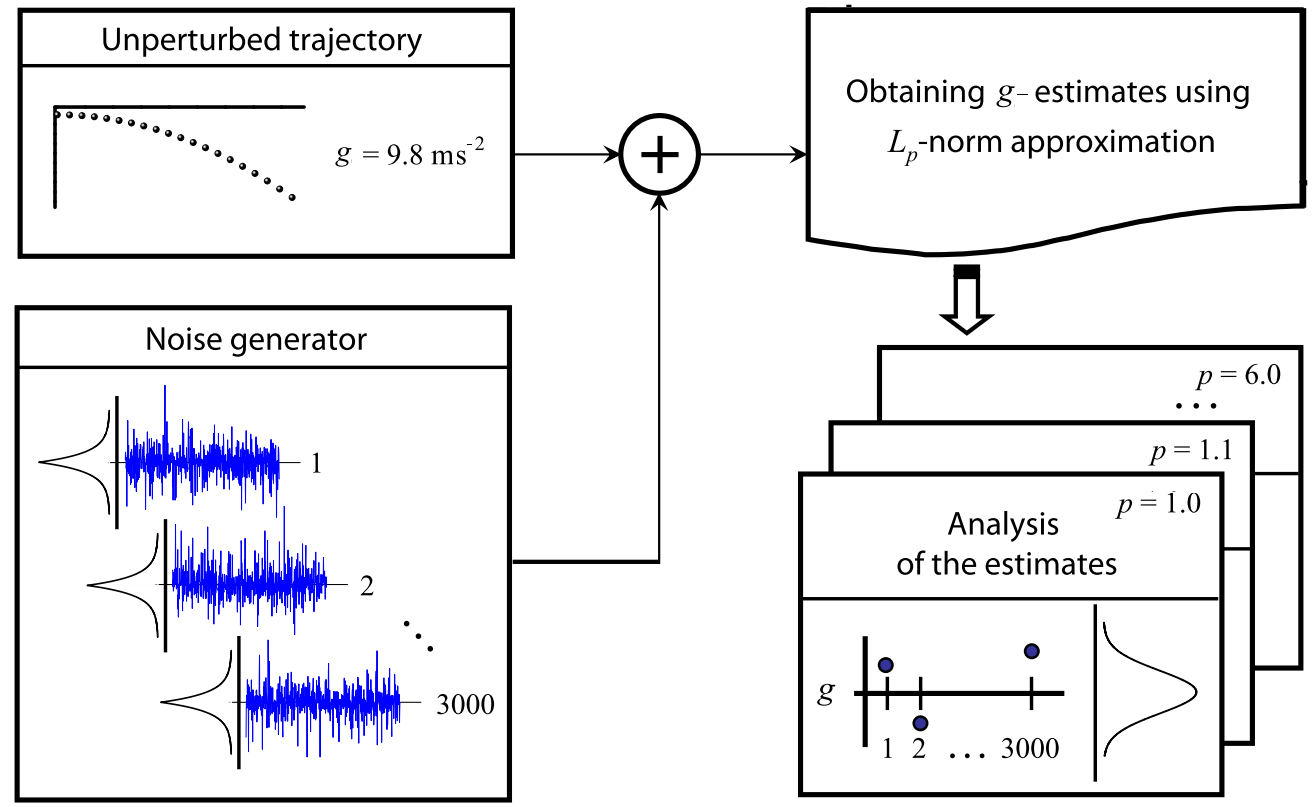}
\caption{Numerical simulation of the properties of the ballistic trajectory approximation in the $L_p$-norm.}
\label{fig_SIMULATION}
\end{figure}

\begin{figure}[ht]
\centering
\includegraphics[height=65mm]{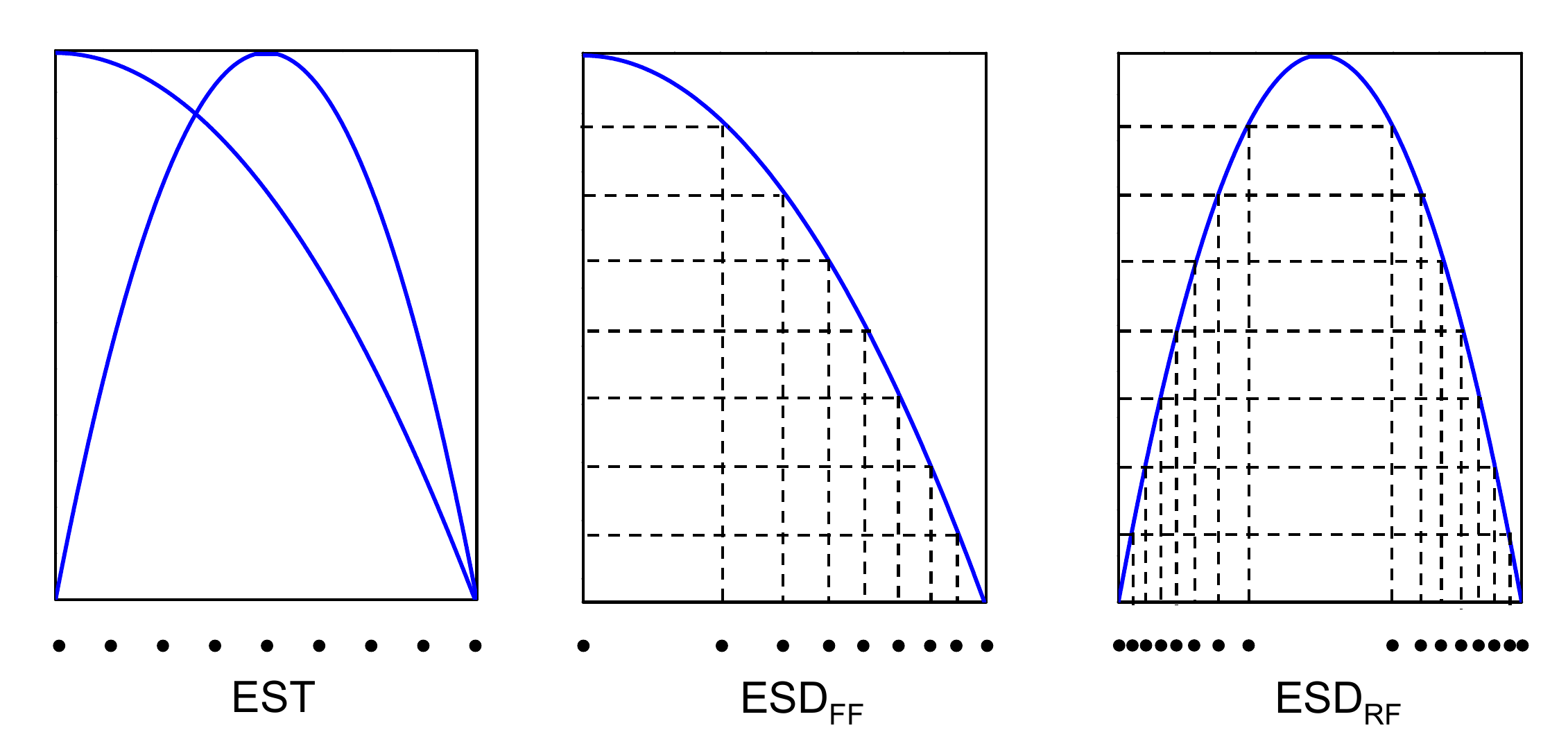}
\caption{Experiment designs used in the simulation:\\
EST -- equally spaced in time. Time intervals do not depend on the trajectory, can be used in both free-fall and rise-and-fall gravimeters;\\
ESD$_{_\textrm{\tiny{FF}}}$ -- equally spaced in distance for the free-fall trajectory;\\
ESD$_{_\textrm{\tiny{RF}}}$ -- equally spaced in distance for the rise-and-fall trajectory.
}
\label{fig_PLANS}
\end{figure}

For the simulations, we added random perturbations to the parabolic trajectory (\ref{eq_z_i}) with known acceleration $g=$ 9.8 ms$^{-2}$\footnote{Other parameters were $z_0$ = 1 nm and $V_0$ = 0 ms$^{-1}$. Zero initial velocity was chosen to maximise the difference between the experiment designs.}, and then recovered the acceleration from the perturbed parabola using the $L_p$-norm approximation (\fref{fig_SIMULATION}). The values of $p$ ran from 1 to 6 with the increment of 0.1, so the standard least-squares solution ($p$=2) was also available from the simulation and was used for comparison. 
To build the parabola, 700 data points $\{T_1, ..., T_{700}\}$ representing 699 time intervals with the  common start at $T_1$ were used, spanning the total time $T_{700}$ = 0.22 s. We considered 3 distributions of the $T_i$-points (experiment designs) found in different types of absolute gravimeters (\fref{fig_PLANS}). The EST (equally spaced in time) design has the uniform distribution of the $T_i$-points and is used in both free-fall and rise-and-fall gravimeters. The ESD$_{_\textrm{\tiny{FF}}}$ design has the minimal density at the start, linearly increasing towards the end of the interval. The ESD$_{_\textrm{\tiny{RF}}}$ design has the minimal density in the centre, linearly increasing towards the start and the end of the interval. These two designs are found in free-fall (FF) and rise-and-fall (RF) gravimeters with the levels equally spaced in distance (ESD).
For every type of perturbation we simulated 3000 drops and, after approximating the trajectory in different $L_p$-norms, analysed the results. Every realisation of the simulated noise was used in all three experiment designs by assigning either EST, ESD$_{_\textrm{\tiny{FF}}}$, or ESD$_{_\textrm{\tiny{RF}}}$ realisations of the vector $\bi{T}=\{T_1, ..., T_{700}\}$ to the same noise vector $\bepsilon=\{\epsilon_1, ... , \epsilon_{700} \}$.

The simulations and data analysis were performed with the MATLAB$^{\textregistered}$ software.
\subsection{Simulation results for the random uncorrelated perturbations}
The trajectory noise in absolute gravimeters comes from a number of sources, such as ground vibrations, laser stabilisation, digital signal acquisition,  fringe counting, etc., so the combined noise most often is of the Gaussian type. Sometimes, however, individual sources can dominate, creating other noise distributions. We did the simulations for several noise types, all having the standard deviation of 1 nm. 

For all noise distributions, the distribution of the $g$-estimates was normal, for any $p$. 
\begin{figure}[ht]
\centering
\includegraphics[height=80mm, angle=90]{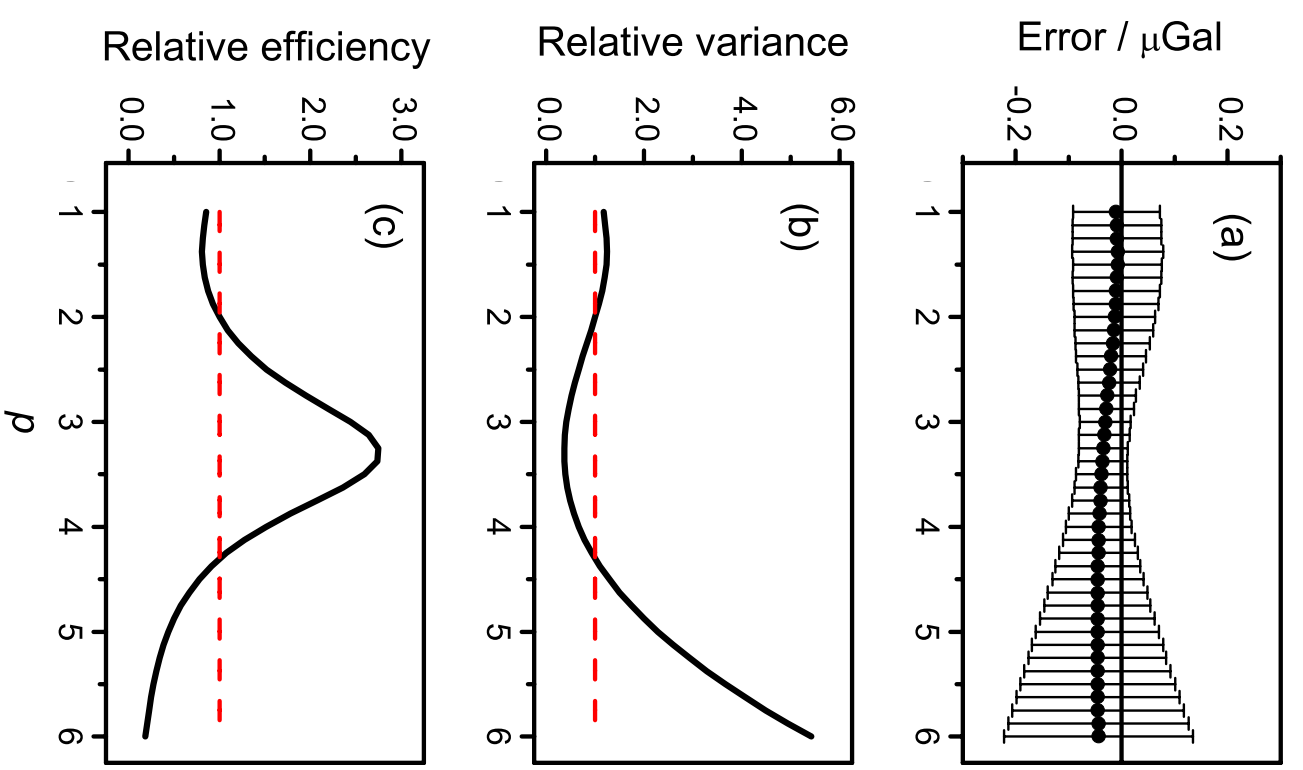}
\caption{Three ways of graphical representation of the simulation results (noise: Uniform, trajectory sampling: EST)\\
a -- mean gravity estimates shown relatively to the known value of $g$ = 9.8 ms$^{-2}$ (1 $\upmu$Gal = $10^{-8}$ ms$^{-2}$) and their standard deviations as error bars, for different $L_p$-norms;\\
b -- variance of the $L_p$-norm estimates relative to the variance of the standard least-squares solution;\\
c -- relative efficiency of the estimates as ratio of the variance of the $L_2$-estimate to the variance of the $L_p$-estimate -- shows the number of times the variance has decreased compared to the standard least-squares.
}
\label{fig_3_PLOTS}
\end{figure}
For every noise distribution there existed an optimal value of $p$ delivering minimal variance of the estimate (\fref{fig_3_PLOTS}a). The optimal $p$ ranged from 1.4 to 3.3 depending on the noise distribution. This result deviates from the theory \cite{nyquist1983} predicting no optimal $p$ for at least the Uniform distribution.

\begin{figure}[ht]
\centering
\includegraphics[height=80mm, angle=90]{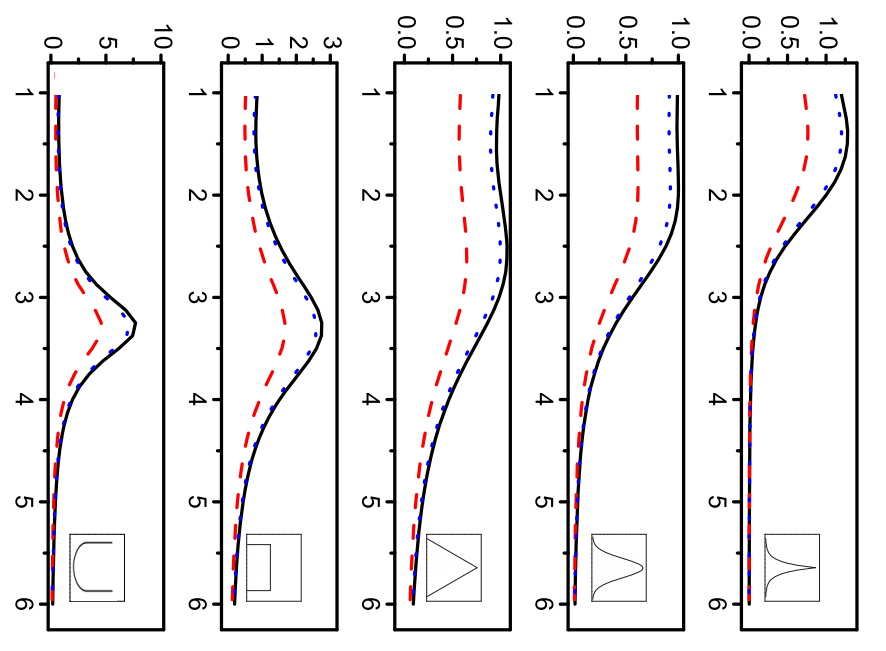}
\caption{
Relative efficiency of the $L_p$-norm estimates (Y-axis) for different noise types. The efficiency is calculated as ratio of the $L_2$-estimate variance for the EST-design to other $L_p$-estimates for all designs.\\
\full EST, \dotted ESD$_{_\textrm{\tiny{RF}}}$, 
\longbroken ESD$_{_\textrm{\tiny{FF}}}$.

}
\label{fig_6_PLOTS}
\end{figure}

For the numeric characteristic of the distribution shape, we used  anitkurtosis\footnote{
The antikurtosis $\upchi$ relates to the kurtosis $\upbeta_2$ as $\upchi=1/\sqrt{\upbeta_2}$ \cite{novitskiy1991}. 
}, because it can only assume values from 0 to 1 for any distribution. We found it more convenient than the traditionally used kurtosis, which can range from 1 to infinity.
The optimal $p$ increased with antikurtosis (\tref{tab_RESULTS}), but not always the variance at the optimal $p$ improved substantially. We visualised the improvement by plotting the variance for every $p$ (\fref{fig_3_PLOTS}b) divided by the variance of the standard least-squares. The factor of the variance improvement is shown on the plot of the inverse ratio (\fref{fig_3_PLOTS}c), which is equivalent to the relative efficiency of the  $L_p$-estimate compared to the standard least-squares one.  
The relative efficiency plots for all simulated noises and experiment designs are presented on the \fref{fig_6_PLOTS}. 
As seen on the figure and at the \tref{tab_RESULTS}, significant improvement of the precision is achieved for the Uniform and Arcsine distributions, if the norm  $L_{3.3}$  is used for the parabola fitting instead of the $L_2$.

As non-linearity of the $L_p$-estimates can entail bias, we did the drop-to-drop comparison of the $L_2$-estimates (theoretically unbiased) with the $L_p$-estimates at the optimal $p$. No statistically significant divergence of the two estimates was detected. We thus confirmed no bias for our approximation procedure, the fact known for the $L_p$-estimates in case of the symmetric noise distributions \cite{money1982, nyquist1983}.
\begin{table}
\caption{Results of the simulation of the absolute gravity estimates by approximating the trajectory in different $L_p$-- norms. The case of random uncorrelated perturbations of the trajectory.}
\begin{indented}
\item[]\begin{tabular}{@{}lccccc}
\br
&Anti-&&\centre{3}{Rel. eff.$^{\rm a}$ for the best $p$}\\
\ns
Noise & kurtosis &  Best &\crule{3}\\
Distribution & $\upchi$ & $p$  & EST & ESD$_{_\textrm{\tiny{FF}}}$ & ESD$_{_\textrm{\tiny{RF}}}$ \\
\mr
Laplace &   0.41 & 1.4 & 1.3 & 1.3 & 1.3\\ 
Normal &   0.58 &  2.0 & 1.0 & 1.0 & 1.0\\ 
Triangle &  0.65 &  2.5 & 1.1 & 1.1 & 1.1\\ 
Uniform &  0.75 &  3.3 & 2.8 & 2.8 & 2.8\\ 
Arcsin &  0.82 & 3.3 & 7.0 & 7.0 & 7.1\\ 
\br
\end{tabular}
\item[] $^{\rm a}$Determined as ratio of the variance of the $L_2$-estimate to the variance of the $L_p$-estimate for the same experiment design.
\end{indented}
\label{tab_RESULTS}
\end{table}
\subsection{Simulation results for the ``harmonic in the Gaussian noise'' perturbations}
According to the \tref{tab_RESULTS}, the $L_p$--estimates with $p>2$ can provide a significant gain in the efficiency when the errors are distributed by the Uniform or Arcsine law. In trajectory tracking such errors are often caused by harmonic perturbations of the reference frame. We modelled the perturbations by the sum of a sinusoid and Gaussian noise. In every simulated drop, a sinusoid with a given frequency $f$ and amplitude of 1.41 nm was assigned an initial phase randomly taken from the uniform distribution on $[0, 2\pi]$, and then sampled according to the experiment design. The sinusoid samples were added by the Gaussian noise, one noise realisation for all three experiment designs. The signal-to-noise ratio (SNR) is defined as quotient of the standard deviations of the sinusoid and the noise, so the infinite SNR means no random noise.

The results of the simulation for the frequencies of 17 Hz, 35 Hz, 55 Hz and the SNR values of $\infty$, 100, 10, 2 are shown in the \tref{tab_RESULTS_SIN}. The antikurtosis $\upchi$ of the perturbation was from 0.70 to 0.82, typical for the range occupied by the Uniform and Arcsine distributions. The resulting $g$--estimates have passed the normality test (Lilliefors , 5\%) only for some low SNR values. The optimal $p$ and relative efficiencies  for which the normality was observed are shaded in the \tref{tab_RESULTS_SIN}. Like for uncorrelated noises (\tref{tab_RESULTS}), there always existed an optimal $p$ delivering the lowest value to the variance of the $L_p$-estimate. The optimal $p$ was between 2.8 and 3.8, yielding the variance decrease of the estimate between 1.2 and 30.6 times. The values of $p$ and the efficiency gain have a complicated dependence on the signal frequency, the experiment design, and the SNR, so the values have to be evaluated experimentally in every practical case.

%
%
\begin{table}
\caption{Results of the simulation of the absolute gravity estimates by approximating the trajectory in different $L_p$-- norms. The trajectory perturbations are ``harmonic in the Gaussian noise.''}
\begin{indented}
\lineup
\item[]\begin{tabular}{@{}ccccccccccc}
\br
&&\centre{3}{EST}&\centre{3}{ESD$_{_\textrm{\tiny{FF}}}$}&\centre{3}{ESD$_{_\textrm{\tiny{RF}}}$}\\
\ns
$f$/Hz & SNR & \crule{3}&\crule{3}&\crule{3}\\
 & & $\upchi$ & $p$ & rel.eff.$^{\rm a}$& $\upchi$ & $p$ & rel.eff.$^{\rm a}$& $\upchi$ & $p$ & rel.eff.$^{\rm a}$\\
\mr
17 & $\infty$ & 0.81 & 3.6 & 5.5 & 0.81 & 3.5 & 17.5 & 0.78 & 3.4 &\01.5\\ 
   & 100      & 0.81 & 3.6 & 5.8 & 0.81 & 3.5 & 16.9 & 0.78 & 3.4 &\01.5\\ 
   & \010     & 0.80 & 3.6 & 5.5 & 0.80 & 3.5 & 17.7 & 0.78 & 3.4 &\01.6\\ 
   & \0\02    & 0.70 &\cc 3.6 &\cc 2.7 & 0.70 &\cc 3.8 &\cc \07.4 & 0.68 & 3.3 &\01.2\\
\\\ns\ns
35 & $\infty$ & 0.82 & 3.5 & 29.3 & 0.82 & 3.4 & 22.2 & 0.82 & 3.6 & 14.9\\ 
   & 100      & 0.82 & 3.5 & 29.5 & 0.82 & 3.4 & 23.2 & 0.82 & 3.5 & 15.1\\ 
   & \010     & 0.81 & 3.5 & 22.0 & 0.81 & 3.4 & 19.6 & 0.81 & 3.6 & 13.7\\ 
   & \0\02    & 0.70 &\cc 3.5 &\cc\02.7 & 0.70 &\cc 3.4 &\cc\03.3 & 0.70 &\cc 3.5 &\cc\03.3 \\
\\\ns\ns
55 & $\infty$ & 0.82 & 3.3 & 16.0 & 0.82 & 3.3 & 28.5  & 0.82 & 3.3 & 25.8\\ 
   & 100      & 0.82 & 3.3 & 15.8 & 0.82 & 3.3 & 28.8 & 0.82 & 3.3 & 25.6\\ 
   & \010     & 0.80 &\cc 3.3 &\cc 15.6 & 0.80 &\cc 3.3 &\cc 30.6 & 0.81 & 3.3 & 25.3\\ 
   & \0\02    & 0.70 &\cc 2.8 &\cc\01.3 & 0.70 &\cc 3.1 &\cc\02.3  & 0.70 &\cc 3.1 &\cc\02.2 \\

\br
\end{tabular}
\item[] $^{\rm a}$Determined as ratio of the variance of the $L_2$-estimate to the variance of the $L_p$-estimate for the same experiment design.
\end{indented}
\label{tab_RESULTS_SIN}
\end{table}
\section{Example of using the $L_p$-estimates on real data}
\label{sec_examples}
\begin{figure}[ht]
\centering
\includegraphics[height=75mm]{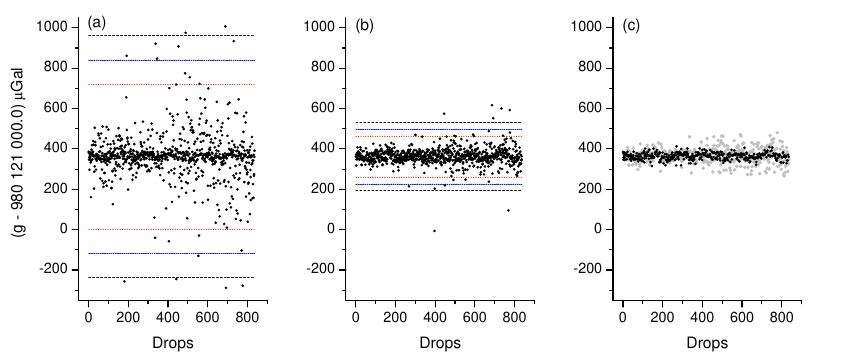}
\caption{Improving precision of the absolute gravity estimates by using the $L_p$-norm approximation of the free-fall trajectory:\\
(a) standard least-squares approximation ($p$=2), the bars are for the 3$\upsigma$, 4$\upsigma$, and 5$\upsigma$; \\
(b) approximation for $p$=3.5, the bars are for the 3$\upsigma$, 4$\upsigma$, and 5$\upsigma$;\\
(c) mixed $p$=2/$p$=3.5 approximation based on the antikurtosis of the residuals: black dots -- Gaussian group ($p$ = 2), grey dots -- Arcsine group ($p$ = 3.5).
}
\label{fig_S_L2}
\end{figure}

The field measurement of absolute gravity are often performed under the harsh geophysical conditions that deteriorate the performance of the instruments. Oftentimes it's not possible to repeat the measurements or improve the conditions, so the researches have to make the best possible gravity estimates using the available data.
As example, we consider a series of measurements taken by the TAG-1 absolute gravimeter developed by the Earthquake Research Institute of the University of Tokyo. The instrument features a laser interferometer with built-in accelerometer for the independent measurement and correction of the ground vibrations, and a homodyne quadrature fringe signal detector \cite{araya2014, svitlov2014}. The free-fall trajectory of about 12 cm is sampled with the frequency of 10 MHz (EST-design). A typical drop-to-drop scatter of the $g$-values at the station ‘Esashi’ is about 20 $\upmu$Gal\footnote{1 $\upmu$Gal = $10^{-8}$ ms$^{-2}$}, so the standard deviation of the mean below 1 $\upmu$Gal is reached in several hundreds drops. The bandwidth of the built-in accelerometer is up to 4 Hz, so the increased intensity of the high-frequency seismic noise could significantly disturb the measurements.
The figure \ref{fig_S_L2}(a) shows one such noisy 
set of gravity estimates taken on December 23, 2011 at the `Esashi' gravity station, after applying standard tidal, atmospheric, and polar motion corrections. To obtain the $L_p$--estimates, the trajectory was decimated to 2 kHz, the standard speed-of-light and gradient corrections were applied to the time and distance coordinates respectively, then the gravity was found using the standard least-squares approximation. 
The set fails the normality test, making the standard `three-sigma' rejection too restrictive for the obtained heavily-tailed distribution of the $L_2$-estimates. However, neither four, nor five sigmas made much difference to the mean $g$-value or its standard deviation, which remained at about 3 $\upmu$Gal (\tref{tab_appxs}).

\begin{figure}[ht]
\centering
\includegraphics[height=60mm]{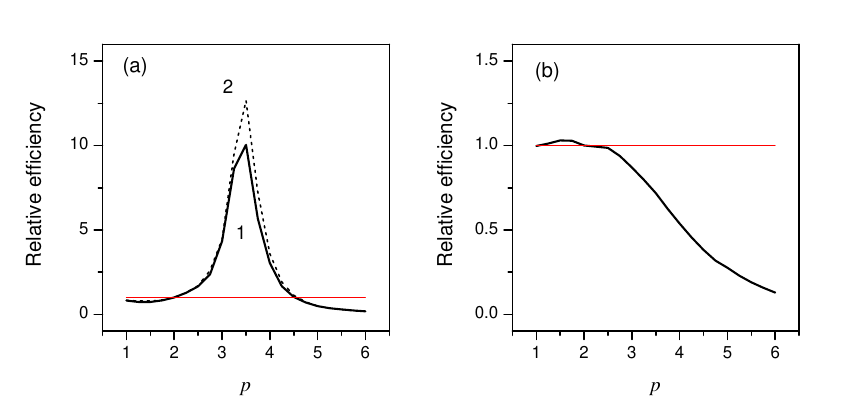}
\caption{Relative efficiency of the gravity estimates with the $L_p$-norm approximation of the trajectory:\\
(a), 1: The entire set with the $5\upsigma$ censoring. The variance of the $L_p$-approximation for $p$=3.5 decreases 10 times.\\
(a), 2: The portion of the set with high antikurtosis values. The variance of the $L_p$-approximation for $p$=3.5 decreases 12.5 times. $3\upsigma$ censoring applied.\\
(b): The Gaussian portion of the set. The standard least-squares ($p=2$) are close to optimal. $3\upsigma$ censoring applied.
}
\label{fig_EX_EFF}
\end{figure}

\begin{figure}[ht]
\centering
\includegraphics[height=100mm]{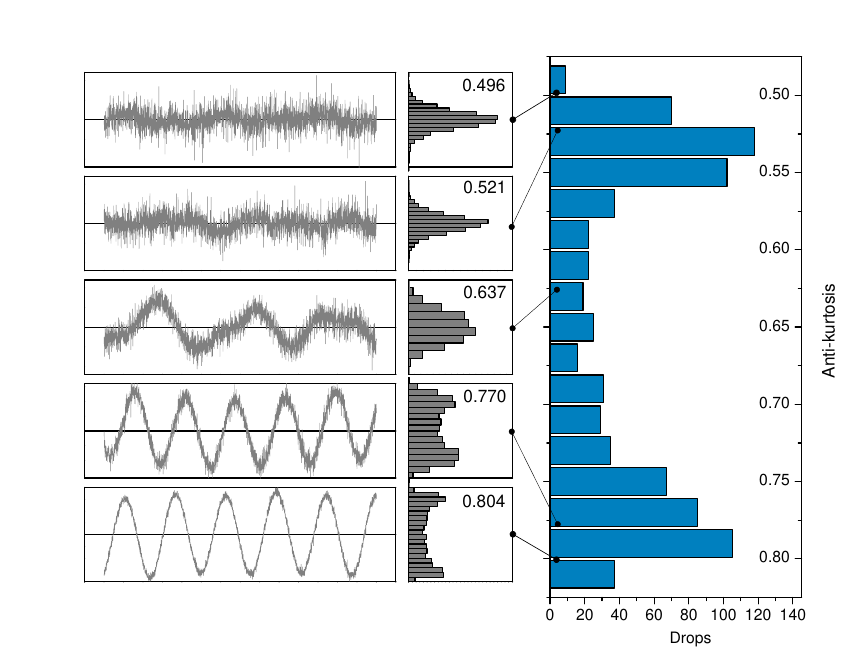}
\caption{Antikurtosis composition of the noise in the data set.\\
The histogram on the right-hand side shows the distribution of the antikurtosis for the noises found in the set's drops. The histograms in the middle and the plots on the left-hand side show examples of the noises present in the set. Due to the inclement seismic conditions, the noises cover a wide range of distributions, from Laplace to Arcsine.   
}
\label{fig_AK_SET}
\end{figure}

We then approximated the trajectory in the $L_p$-norm for $p$ ranging from 1 to 6 (see section \ref{sec_IRLS}). For each $p$ we obtained the variance of the set gravity estimate and plotted the relevant relative efficiency (\fref{fig_EX_EFF}). According to the plot (\fref{fig_EX_EFF}(a), curve~1), the variance at $p$=3.5 decreased several times w.r.t. $p$=2, reducing the standard deviation of the mean to about 1 $\upmu$Gal (\tref{tab_appxs}).
Based on the simulations (\sref{sec_simulation}), the peak of the efficiency at $p>2$ is suggestive of the residuals with non-Gaussian distributions. We investigated the residuals and found  the values of antikurtosis ranging from 0.487 to 0.813, which corresponds to the wide range of distributions -- from Laplace to Arcsine (\fref{fig_AK_SET}).
The likely reason for the non-Gaussian errors was resonant oscillations of the reference mirror caused by the excessive seismic noise.
The most prevalent in the set were residuals with the Gaussian and Arcsine distributions. According to the simulations (\sref{sec_simulation}), the most efficient $L_p$-approximations for the Arcsine-noised trajectory were achieved for the values of $p$ between 3 and 4, which explains the decrease of the uncertainty. However, for the Gaussian noises the approximation with $p$ in that range can be several times less efficient than the standard least-squares (\fref{fig_6_PLOTS}, second from the top). This led us to the idea of splitting the set into two parts with predominantly Gaussian and Arcsine noises, and independent optimising the $L_p$-norm on each set. We have split the groups at the antikurtosis $\upchi$  = 0.65, about in the middle of two peaks at the \fref{fig_AK_SET}. Gravity estimates in each group have passed the normality test (Lilliefors , 5\%), so further processing was carried out with the $3\upsigma$ rejection, each group with its own $\upsigma$.  The splitting has increased the efficiency of the estimate for the Arcsine group with $p$=3.5 from 10.0 to 12.5 (\fref{fig_EX_EFF}(a), 2). For the Gaussian group we used the standard least-squares with  $p$=2.0, which helped to avoid the decrease of the efficiency at $p$=3.5 for that group (\fref{fig_S_L2}(b)). Combining the estimates for both groups has reduced the standard deviation of the weighted mean to 0.76 $\upmu$Gal (\fref{fig_S_L2}(c)).
The gravity estimates obtained with different approximations are summarised in the \tref{tab_appxs}.
\begin{table}
\caption{Estimates of the absolute gravity value obtained with different modes of the approximation of the free-fall trajectory }
\begin{tabular}{@{}lccccc}
\br
Mode of approximation & Rejection & Accepted & Set mean & Set std.dev.& Mean's std.dev.\\
of the trajectory & limits & drops & $\bar{g}$ / $\upmu$Gal & $\bar{\upsigma}$ / $\upmu$Gal & $\bar{\upsigma}$ / $\upmu$Gal \\
\mr
Whole set, $L_2$ & $ 3\upsigma$ & 784 & 980 121 363.52 & 80.44  & 2.87  \\
                 & $ 4\upsigma$ & 803 & 980 121 362.38 & 95.24 & 3.36  \\
                 & $ 5\upsigma$ & 811 & 980 121 363.96 & 107.31 & 3.77  \\
\\\ns\ns
Whole set, $L_{3.5}$ & $ 3\upsigma$ & 802 & 980 121 362.70 &  29.51 & 1.04 \\
                     & $ 4\upsigma$ & 817 & 980 121 363.31 & 32.67 & 1.14 \\
                     & $ 5\upsigma$ & 820 & 980 121 362.76 & 33.88 & 1.18 \\
\\\ns\ns
Split groups, $L_2$/$L_{3.5}$ 
& $ 3\upsigma/3\upsigma$ & 396/406 & 980 121 364.50$^{\rm a}$ & 16.37/38.18   & 0.76$^{\rm b}$  \\ 
\br
\end{tabular}
$^{\rm a}$ Weighted average of two groups, the weights are the reciprocals of the groups' variances\\
$^{\rm b}$ Standard deviation of the weighted average

\label{tab_appxs}
\end{table}

\section{Conclusions}
\label{sec_conclusions}
We have simulated the approximation of the ballistic trajectory in different $L_p$- norms, in view of using it to improve the absolute gravity estimates. To perform the approximation, the Iteratively Re-weighted Least Squares method was implemented. The estimates were simulated for three different plans of the experiment corresponding to two types of the ballistic trajectory (free-fall, rise-and-fall) and two schemes of data location (equally spaced in time or in distance). Each plan was applied to several perturbation types, including the cases of random uncorrelated noises and a harmonic in the Gaussian noise. As an application example, the simulation results were used to improve the absolute gravity estimates obtained at the `Esashi' gravity station by the TAG-1 gravimeter.

The following conclusions can be drawn from the study.
\begin{enumerate}
\item The Iteratively Re-weighted least-squares with two iterations approximate the ballistic trajectory in the $L_p$-norm with the accuracy sufficient for the purposes of absolute gravimetry. No statistically significant bias of the estimate was found for the wide range of the trajectory noises.
\item The $L_p$-norm approximation of the ballistic trajectory with $p>2$ estimates gravity more precisely than the standard least-squares ($p$=2), if the trajectory perturbations deviate from the Gaussian towards more platykurtic distributions, i.e. those with flattish or concave shape.
\item The observed efficiency maximum for certain values of $p$ on platykurtic distributions disagrees with the prediction of the work \cite{nyquist1983}, where the monotonous increase of the efficiency with the norm power $p$ is obtained for the Uniform distribution. This could possibly be explained by a limited iterations of our IRLS procedure, resulting in a ``quasi'' rather than the ``exact'' $L_p$-approximation.
\item For the same measurement interval $T$ and fixed number of data points $N$, the EST sampling of the free-fall trajectory makes the gravity estimates about twice as efficient as those with the data equally spaced in distance, for any $p$. This points to an obvious way of improving the modern free-fall absolute gravimeters, which predominantly use the ESD scheme. For the symmetric rise-and-fall trajectory, the efficiency of the EST and ESD sampling is about the same. 
%
    %
%
%
\end{enumerate}
The presented study is a first attempt to obtain the gravity estimates by approximating the ballistic trajectory in the $L_p$-norm with $p\ne 2$. Many important questions are yet to be answered by future work. The questions include, but are not limited to
\begin{enumerate}
\item The scope of the applicability of the $L_p$-norm approximation. For the purposes of absolute gravimetry we investigated only the estimates of the quadratic coefficient of the second-order polynomial model. The results can be applied to other models to build more efficient estimates in wider range of applications.
%
%
%
%
\item Auto-correlation of the residuals. The main source of the platykurtic distributions is the low-frequency noises. Our simulations of the correlated Arcsine noises and processing of the real data suggest that the efficiency of the $L_p$-approximation is defined mostly by the noise distribution and is not deteriorated by the auto-correlation of the residuals. Still, the influence of the auto-correlation calls for more thorough investigation, including the comparison to other algorithms of the low-frequency noise filtering.
\item Contaminated distributions within a drop.  Further investigations are desirable to better understand the properties of the $L_p$-norm approximation for the mixed noise distributions, such as harmonics in the Gaussian noise. 
\item Contaminated distributions of the set and set splitting. When a set has drops with different noise types, a single value of $p$ may not yield the best $L_p$-approximations for all drops. In the example (\sref{sec_examples}) we have split the set in two at the centre of the antikurtosis histogram. Though we gained certain uncertainty decrease, there may exist better ways of set splitting, worth to be investigated.
\item The reasons of the departure of the simulated results from those predicted in \cite{nyquist1983} have to be investigated to better understand the properties of the IRLS algorithm of the $L_p$-approximation.
%
%
\end{enumerate}
%
%
%
%

\section*{References}


\end{document}